\documentclass[12pt]{iopart} 
\usepackage{graphicx}
\newcommand{\sauron}{\texttt{SAURON}}
\begin{document}

\title[Supermassive Black Holes with Integral-Field Spectroscopy]{The Nuclear Orbital Distribution in Galaxies as a Fossil Record of Black Hole Formation from Integral-Field Spectroscopy}

\author{Michele Cappellari and Richard M.\ McDermid}

\address{Leiden Observatory, Postbus 9513, 2300 RA Leiden, The Netherlands}

\ead{cappellari@strw.leidenuniv.nl}

\begin{abstract}
In the past decade, most effort in the study of supermassive black holes (BHs) has been devoted to measuring their masses. This lead to the finding of the tight $M_{\rm BH} - \sigma$ relation, which indicates the existence of strong links between the formation of the BH and of their host spheroids. Many scenarios have been proposed to explain this relation, and all agree on the key role of BHs' growth and feedback in shaping their host galaxies. However the currently available observational constraints, essentially BH masses and galaxy photometry, are not sufficient to conclusively select among the alternatives. A crucial piece of information of the black hole formation is recorded in the orbital distribution of the stars, which can only be extracted from high-resolution integral-field (IF) stellar kinematics. The introduction of IF spectrographs with adaptive optics on large telescopes opens a new era in the study of BHs by finally allowing this key element to be uncovered. This information will be complementary to what will be provided by the LISA gravitational wave satellite, which can directly detect coalescing BHs. Here an example is presented for the recovery of the orbital distribution in the center of the giant elliptical galaxy M87, which has a well resolved BH sphere of influence, using \sauron\ IF kinematics.
\end{abstract}

\submitto{\CQG {\bf 22}, (2005) S347-S353}

\section{Introduction}

\begin{figure}
\centering
\includegraphics[width=\columnwidth]{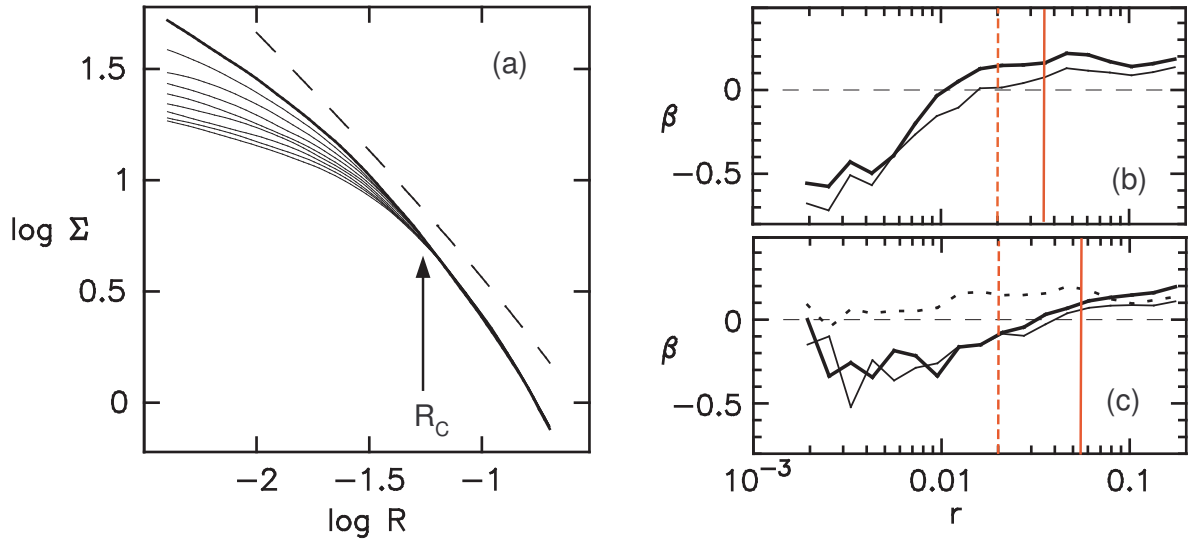}
\caption{Observable signatures of the formation of binary BHs in galaxies, predicted by $N$-body simulations. (a) projected surface brightness. From top to bottom the different lines indicate the evolution, starting from the time $t_h$ of the formation of the hard black hole binary. The vertical arrow indicates the location of the core radius $R_c$ for the lowest profile. (b) orbital distribution (anisotropy; here defined as $\beta=1-\sigma_t^2/\sigma_r^2$, where $\sigma_r$ is the radial and $\sigma_t$ the tangential component of the velocity dispersion tensor) at the time $t_h$. The thick and thin lines are measured along the major and minor axis respectively. The dashed vertical line indicates the radius $R_{\rm BH}$ of the BH sphere of influence, while the solid vertical line corresponds to the radius $R_c$. (c) same as in the previous panel, but at the end of the simulation. Inside the radius where the density profile is `scoured' by the BH binary, the orbital distribution becomes tangentially biased. For comparison the dotted line shows the anisotropy for a simulation with a single BH, where the orbital distribution remains mildly radially anisotropic. (Adapted from Milosavljevi\'c \& Merritt 2001).}
\end{figure}

In recent years there has been tremendous progress, primarily from space-based observations, in our understanding of the distribution of central black hole (BH) masses and the relation between BHs and their host galaxies (see de Zeeuw 2004 for a review). The resulting picture of BH demography is summarized by the correlation between BH mass and absolute spheroid luminosity (e.g., Magorrian et al.\ 1998) and the much tighter correlation between BH mass and galaxy central velocity dispersion (the $M_{\rm BH}-\sigma$ relation; Ferrarese \& Merrit, 2000; Gebhardt et al.\ 2000). It is now clear that BHs are nearly ubiquitous in galaxies, and there is increasing evidence that BH formation plays a key role in driving galaxy formation, via feedback processes. The correlations obtained from nearby galaxies are also being extended to study the high-redshift universe. To date, however, the only observable constraints on the BH formation process are the BH masses and the nuclear stellar density profiles of nearby galaxies. Until now, very little is known about the nuclear orbital distribution of the stars, which constitutes a fossil record of the BH formation process. The advent of high spatial resolution {\em integral-field} spectroscopic observations is opening a new era in the study of BHs by allowing this crucial piece of information to be reliably extracted.

\section{Observable Constraints on Black Hole Formation Scenarios}

Spheroidal components of galaxies are thought to form through a series of hierarchical mergers of smaller systems, which themselves have central BHs. As two systems merge, their central BHs rapidly fall to the minimum of the potential well due to dynamical friction, and create a binary system at the center of the remnant. The binary looses energy and shrinks by capturing stars on radial orbits which pass nearby, and ejecting them at much higher velocities, thus dynamically heating the central regions and lowering its stellar density, scouring out a `core' of radius $R_c$ in the photometric profile (Quinlan \& Hernquist 1997; Milosavljevi\'c \& Merritt 2001). In this way, a coalescing BH binary can eject a total mass of stars similar in order to its own mass, and influence the orbital structure of the remnant nucleus out to the radius $R_c$, which is several times larger than the actual BH radius of influence. The imprint of this process is found in the central orbital structure, which can be quantified e.g.\ in terms of the ratio of radial to tangential orbits (Fig.~1). BHs can also grow by acquiring gas which sinks via dissipation to the bottom of the galaxy potential well. In this case, simulations show that a cusp forms in the density profile when the gas subsequently forms stars, while the orbital distribution is only weakly affected. These two black hole formation mechanisms have been proposed as an explanation for the dichotomy of central luminosity profiles of early-type galaxies (Faber et al.\ 1997). While the details of the theoretical predictions for the effects of the BH growth on the orbital distribution are still under lively debate (Merritt \& Milosavljevi\'c 2004; Makino \& Funato 2004), due to the difficulty of constructing accurate $N$-body simulations with very large numbers of stars near the BH singularity, it already appears clear that BHs have a profound influence on the structure of galaxy nuclei, which retain the fossil record of the formation process. One of the strongest observable constraints on the BH formation scenarios is its effect on the central orbital structure.

\begin{figure}
\includegraphics[width=\columnwidth]{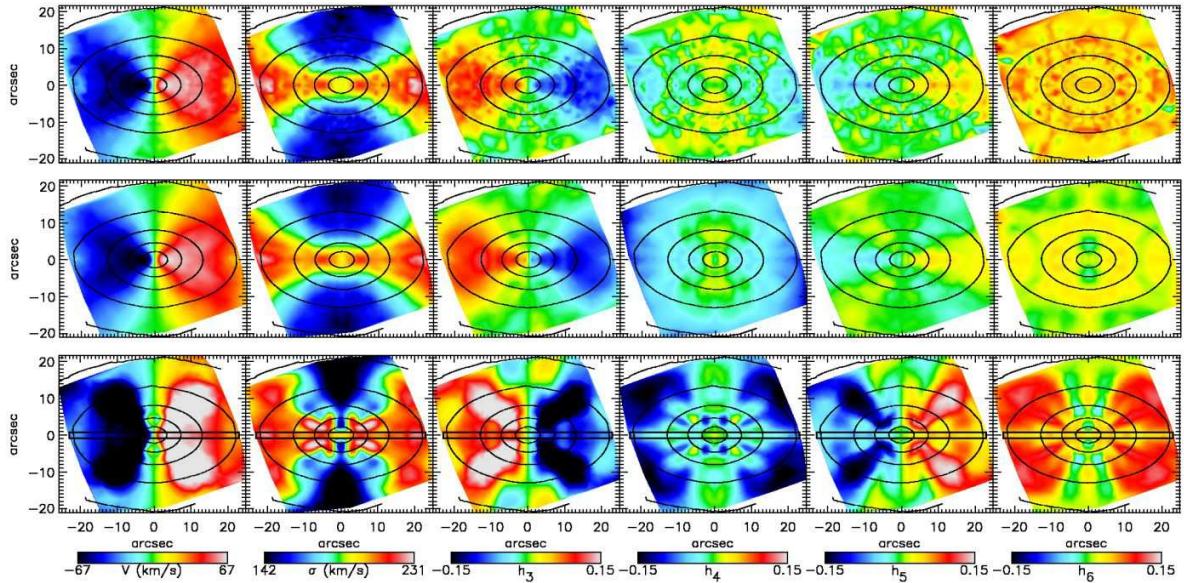}
\caption{Each column from left to right shows the kinematic moments: mean velocity, $\sigma$, and higher Gauss-Hermite moments h$_3$--h$_6$ (see text). {\it Top row:} 2D-binned, symmetrized and linearly interpolated {\tt SAURON} observations of the early-type galaxy NGC\,4473 from Emsellem et al.\ (2004). {\it Middle row:} Three-integral Schwarzschild model fit using the full {\tt SAURON} field. The analysis of this model shows that the galaxy contains two flattened counterrotating spheroids, of unequal mass (Cappellari et al.\ 2004). This explains the unusual high velocity dispersion along the major axis and the V-shaped velocity field. {\it Bottom row:} The same model, but fitted only to a simulated major-axis long slit extracted from the {\tt SAURON} field. All models are regularized and have a smooth phase space. The kinematics are fitted perfectly along the slit, and the galaxy density is everywhere well reproduced; outside the slit, however, the model is completely different from the true galaxy. Only using the full 2D field are the kinematics, and therefore the orbital structure, properly constrained.}
\end{figure}

\section{Limitations of high-resolution long-slit studies}

Probing the BH radius of influence demands sub-arcsecond spatial resolution, even for nearby galaxies. For this reason, {\tt STIS} onboard HST has been used for many well-determined BH masses measured from dynamical modeling of the kinematics of stars or gas. There are also several studies currently in progress to establish black hole masses using adaptive optics (AO)-assisted long-slit measurements from the ground, using {\tt NAOS-CONICA} (VLT) and {\tt NIRI-Altair} (Gemini). There are strong limitations, however, to what can be achieved by long-slit studies.  A single slit observation is not sufficient to detect possible signs of non-axisymmetry in the kinematics, resulting from a nuclear bar, for example. Therefore the symmetry assumptions generally made in the dynamical models, used to measure BH masses, cannot be tested.

More importantly, the orbital distribution (e.g., the anisotropy) cannot be recovered without knowledge of the line-of-sight velocity-distribution (LOSVD)  of the stars at {\em all} spatial positions on the galaxy image projected on the sky (Fig.\ 2). Integral-field kinematic data is therefore the only way to reliably constrain the orbital structure (Cappellari et al.\ 2004), and general orbit-based methods can then be used in realistic conditions, for a given potential, to extract the orbital distribution from the two-dimensional (2D) kinematics (Krajnovi\'c et al.\ 2005). Ignorance of the anisotropy due to only having a single long-slit directly translates into large uncertainties in the BH mass determination (Verolme et al.\ 2002), as well as invalidating any constraint on the BH formation scenario.  Although observational studies of the velocity anisotropy around BHs exist (Gebhardt et al.\ 2003), these are based on single-slit information, from HST/{\tt STIS}, and so cannot place tight constraints on the orbital structure.  Avoiding these issues by obtaining multiple slit positions has a prohibitive observational expense, and no such studies exist.

Until now, no instrument could provide integral-field kinematics at the resolution of HST (but see Bacon et al. 2001a), so it was impossible to reliably infer the orbital distribution near the nuclear black holes. The advent of AO-fed IF units on large telescopes, like {\tt SINFONI} (VLT), {\tt OASIS} (WHT), {\tt NIFS} (coming on Gemini) and {\tt OSIRIS} (coming on the Keck telescopes), is finally going to change this situation. In particular, the future prospect of laser guide star-based AO systems on these instruments will allow statistically meaningful samples to be obtained. Observations from such instruments, combined with general dynamical models, will provide the first reliable determination of the effect of a black hole on the nuclear stellar orbits, giving a unique insight into the formation of BHs and their host systems.

\section{Nuclear orbital distribution of M87 from integral-field kinematics}

\begin{figure}
\includegraphics[width=\columnwidth]{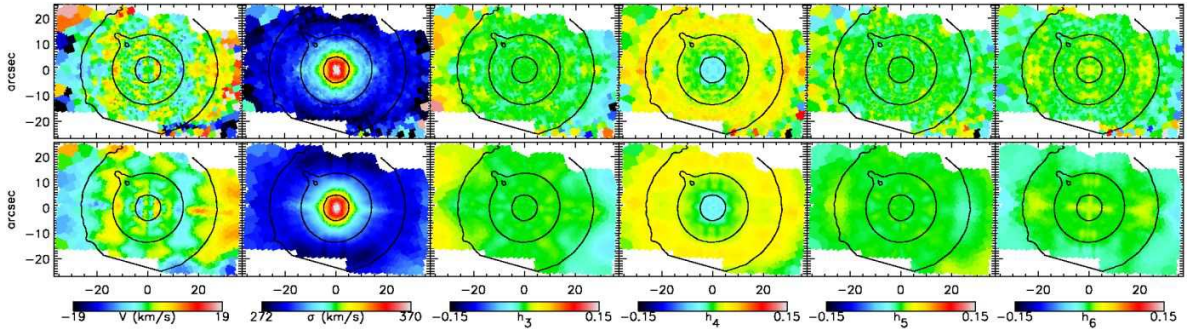}
\caption{Each column from left to right shows the kinematic moments: mean velocity, $\sigma$, and higher Gauss-Hermite moments h$_3$--h$_6$. {\it Top row:} Voronoi 2D-binned and symmetrized {\tt SAURON} observations of the giant elliptical galaxy M87. {\it Bottom row:} Three-integral Schwarzschild model fit to the above kinematics.}
\end{figure}

To demonstrate what IF observations can provide, we present here the recovery of the orbital distribution from general axisymmetric three-integral dynamical modeling of the IF stellar kinematics of M87, which is part of the \sauron\ survey (Bacon et al.\ 2001b; de Zeeuw et al.\ 2002). This galaxy is an ideal candidate for this ground-based study as its BH sphere of influence $R_{\rm BH}\approx1.4''$ is resolved even in natural seeing. Moreover, the core, whose structure is thought to have been shaped by the BH binary, has a radius $R_c\approx7.5''$ (Faber et al.\ 1997) and is very well sampled by the observed \sauron\ kinematics.

The observations were presented in Emsellem et al.\ (2004) and we refer to that paper for details. In brief, the \sauron\ IF observations were spatially binned to a minimum $S/N\approx60$ using the Voronoi 2D-binning algorithm by Cappellari \& Copin (2003) and the stellar kinematics were subsequently extracted with the penalized pixel-fitting (pPXF) method of Cappellari \& Emsellem (2004). This provided, for each Voronoi bin, the mean velocity $V$, the velocity dispersion $\sigma$ and the higher order Gauss-Hermite moments of the velocity, including $h_3$--$h_6$ (van der Marel \& Franx 1993; Gerhard\ 1993). For the three-integral dynamical modeling, we adopted Schwarzschild's (1979) numerical orbit-superposition method, which can fit all kinematical and photometric observations (Rix et al.\ 1997; van der Marel et al.\ 1998; Cappellari et al.\ 2002). A similar approach was adopted by other groups to measure the black hole (BH) masses in galaxy nuclei (e.g. Gebhardt et al.\ 2003; Valluri et al.\ 2004).

\begin{figure}
\centering
\includegraphics[width=0.7\columnwidth]{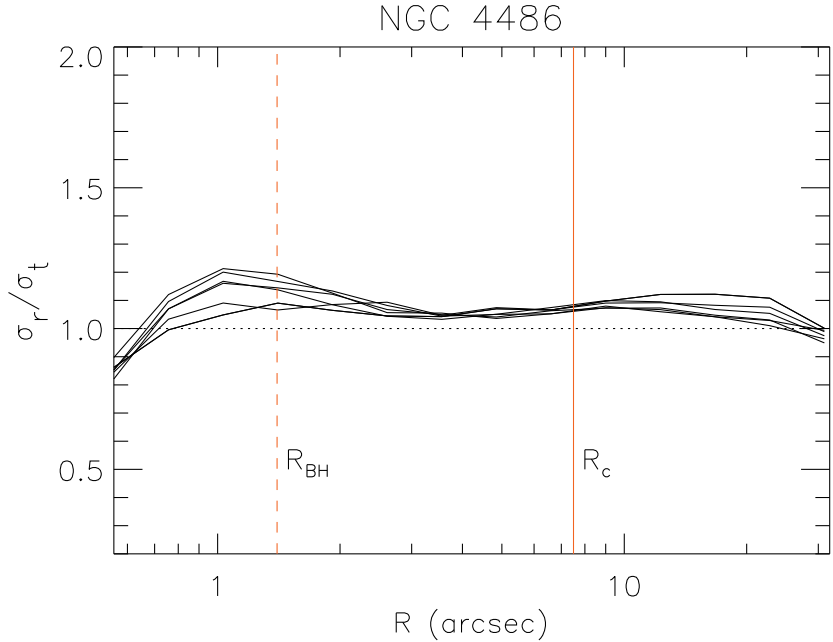}
\caption{Radial profile of the orbital anisotropy $\sigma_r/\sigma_t$ in M87. The different solid lines are measured along different polar angles, in the galaxy meridional plane, and provide an indication of the model uncertainties. The dotted horizontal line corresponds to an isotropic model. The dashed vertical line indicates the radius $R_{\rm BH}$ of the BH sphere of influence, while the solid vertical line corresponds to the location of the break radius $R_c$, which is indicative of the core size. The main galaxy body is mildly radially anisotropic until well inside the sphere of influence of the BH (at the limit of our spatial resolution), where the orbits become tangentially anisotropic.}
\end{figure}

Our best fitting model with constant stellar mass-to-light ratio $M/L$ (Fig.~3; for an assumed inclination $i=90^\circ$) is able to reproduce all the details of the observed kinematics. As our kinematic measurements do not have enough spatial resolution to provide tight constraints on the BH mass, we {\em assume} $M_{\rm BH}\approx3\times10^9 M_\odot$ as derived from the HST gas kinematics (Harms et al.\ 1994; Macchetto et al.\ 1997). The anisotropy profile of the model is presented in Fig.~4. It shows that the main galaxy body is mildly radially anisotropic even inside $R_c$, and it is not until well inside $R_{\rm BH}$ (at the limit of our spatial resolution) that it becomes tangentially anisotropic. This result is consistent with previous models by Dressler \& Richstone (1990), van der Marel (1994) and Merritt \& Oh (1997) which only made use of the velocity dispersion along a single slit position. This is not surprising given the near circular symmetry of the projected observables of this galaxy in the central regions. In a nonrotating spherical stellar system, if the potential is given, the anisotropy is uniquely specified by the velocity dispersion alone (Binney \& Manon 1982). Therefore, the fact that our model is able to fit the higher order moments of the velocity as well, gives us confidence that the orbital distribution is being correctly recovered.

Although little can be learned from a single object and a specific simulation, it is still interesting to see how the observations and the predictions compare. We find that the observed distribution of M87, which is the prototypical giant core elliptical galaxy, is not obviously explained by current simulations of the BH binary core-scouring mechanism. In fact, although the observed anisotropy is consistent with the predicted one at the time $t_h$ of the formation of the hard BH binary (Fig.~1b), the steep surface brightness profile (Fig.~1a) is inconsistent with the observations (Fig.~7 of van der Marel et al.\ 1994). The observed surface brightness profile is better reproduced by the simulations at later times, when the binary had time to scour a deeper core, but the anisotropy profile at that time (Fig.~1c) is inconsistent with the observed one. This discrepancy may be due to the effect of multiple merger events (Milosavljevi\'c \& Merritt 2001), making the simulations and the observations difficult to compare in detail. This practical example demonstrates the method, and highlights the need to perform this kind of study for larger samples of objects, in order to tightly constrain the BH formation mechanism of these massive ($M_{\rm BH}$$>$ $10^6$) BHs in nearby galaxies, and uncover fossil signatures of BHs binaries, which can be compared with detailed predictions from simulations. This information will be complementary to what will be provided by the LISA gravitational wave satellite, which can directly detect coalescing BHs (if this happens) in the smaller mass range $M_{\rm BH}$$<$$10^6$ M$_\odot$, and in a much larger redshift range (up to $z$$\sim$10). Understanding the BHs growth is expected to shed light on the feedback processes by which BHs shape galaxy evolution.

\ack

We thank the \sauron\ team for allowing use of the kinematics of M87, for comments on this paper and for many fruitful discussions. We thank the anonymous referee for useful comments.

\References

\item[] Bacon, R., Emsellem, E., Combes, F., Copin, Y., Monnet, G., \& Martin, P.\ 2001a, A\&A, 371, 409

\item[] Bacon, R., et al.\ 2001b, MNRAS, 326, 23

\item[] Binney, J., \& Mamon, G.~A.\ 1982, MNRAS, 200, 361

\item[] Cappellari, M., \& Copin, Y.\ 2003, MNRAS, 342, 345

\item[] Cappellari, M., \& Emsellem, E.\ 2004, PASP, 116, 138

\item[] Cappellari, M., et al.\ 2002, ApJ, 578, 787

\item [] Cappellari, M., et al.\ 2004, in Carnegie Observatories Astrophysics Series, Vol. 1: Coevolution of Black Holes and Galaxies, ed.\ L.\ C.\ Ho (Pasadena: Carnegie Observatories) (astro-ph/0302274)

\item[] de Zeeuw, P.~T.\ 2004, in Ho, L. C., ed, Carnegie Observatories
Astrophysics Series, Vol. 1: Coevolution of Black Holes and Galaxies, (Cambridge: Cambridge Univ. Press), p. 461

\item[] de Zeeuw, P.~T., et al.\ 2002, MNRAS, 329, 513

\item[] Dressler, A., \& Richstone, D.~O.\ 1990, ApJ, 348, 120

\item[] Emsellem, E., et al.\ 2004, MNRAS, 352, 721

\item[] Faber, S.~M., et al.\ 1997, AJ, 114, 1771

\item[] Ferrarese, L., Merritt, D.\ 2000, ApJ,539, L9

\item[] Gebhardt, K., et al.\ 2000, ApJ, 539, L13

\item[] Gebhardt, K., et al.\ 2003, ApJ, 583, 92

\item[] Gerhard, O.~E.\ 1993, MNRAS, 265, 213

\item[] Harms, R.~J., et al.\ 1994, ApJL, 435, L35

\item[] Krajnovi\'c, D., Cappellari, M., Emsellem, E., McDermid, R., de
Zeeuw, P.T.\ 2005, MNRAS, 357, 1113

\item[] Macchetto, F., Marconi, A., Axon, D.~J., Capetti, A., Sparks,
W., \& Crane, P.\ 1997, ApJ, 489, 579

\item[] Magorrian, J., et al.\ 1998, AJ, 115, 2285

\item[] Makino, J., \& Funato, Y.\ 2004, ApJ, 602, 93

\item[] Merritt, D., \& Oh, S.~P.\ 1997, AJ, 113, 1279

\item[] Merritt, D., \& Milosavljevi\'c, M.\ 2004, in Living Reviews in
Relativity, in press (astro-ph/0410364)

\item[] Milosavljevi{\'c}, M., \& Merritt, D.\ 2001, ApJ, 563, 34

\item[] Quinlan, G.~D., \& Hernquist, L.\ 1997, New Astronomy, 2, 533

\item[] Rix, H.-W., de Zeeuw, P.~T., Cretton, N., van der Marel, R.~
P., Carollo, C.~M.\ 1997, ApJ, 488, 702

\item[] Schwarzschild, M.\ 1979, ApJ, 232, 236

\item[] van der Marel, R.~P., \& Franx, M.\ 1993, ApJ, 407, 525

\item[] van der Marel, R.~P.\ 1994, MNRAS, 270, 271

\item[] van der Marel, R.~P., Cretton, N., de Zeeuw, P.~T., Rix, H.-W.\
1998, ApJ, 493, 613

\item[] Valluri, M., Merritt, D., \& Emsellem, E.\ 2004, ApJ, 602, 66

\item[] Verolme, E.~K., et al.\ 2002, MNRAS, 335, 517

\endrefs

\end{document}